\documentstyle[epsf]{mn}

\title[{\sl XMM-Newton} spectral studies of DP Leo and WW Hor]{First 
{\sl XMM-Newton} observations of strongly magnetic cataclysmic
variables I: spectral studies of DP Leo and 
WW Hor}

\author[G. Ramsay et al.]{Gavin Ramsay$^{1}$, Mark Cropper$^{1}$,
France C\'{o}rdova$^{2}$, 
Keith Mason$^{1}$, Rudi Much$^{3}$, \and Dirk Pandel$^{2}$, Robert Shirey$^{2}$\\
$^{1}$Mullard Space Science Lab, University College London,
Holmbury St. Mary, Dorking, Surrey, RH5 6NT, UK\\
$^{2}$Department of Physics, University of California, Santa
Barbara, California 93106, USA\\
$^{3}$Astrophysics Division, ESTEC, 2200, AG Noordwijk, The
Netherlands}
\date{Received: }

\begin{document}
\outer\def\gtae {$\buildrel {\lower3pt\hbox{$>$}} \over 
{\lower2pt\hbox{$\sim$}} $}
\outer\def\ltae {$\buildrel {\lower3pt\hbox{$<$}} \over 
{\lower2pt\hbox{$\sim$}} $}
\newcommand{\ergscm} {ergs s$^{-1}$ cm$^{-2}$}
\newcommand{\ergss} {ergs s$^{-1}$}
\newcommand{\ergsd} {ergs s$^{-1}$ $d^{2}_{100}$}
\newcommand{\pcmsq} {cm$^{-2}$}
\newcommand{\ros} {\sl ROSAT}
\newcommand{\exo} {\sl EXOSAT}
\def\rchi{{${\chi}_{\nu}^{2}$}}
\newcommand{\Msun} {$M_{\odot}$}
\newcommand{\Mwd} {$M_{wd}$}
\def\Mdot{\hbox{$\dot M$}}
\def\mdot{\hbox{$\dot m$}}

\maketitle

\begin{abstract}
We present an analysis of the X-ray spectra of two strongly magnetic
cataclysmic variables, DP Leo and WW Hor, made using {\sl
XMM-Newton}. Both systems were in intermediate levels of
accretion. Hard optically thin X-ray emission from the shocked
accreting gas was detected from both systems, while a soft blackbody
X-ray component from the heated surface was detected only in DP Leo.
We suggest that the lack of a soft X-ray component in WW Hor is due to
the fact that the accretion area is larger than in previous
observations with a resulting lower temperature for the re-processed
hard X-rays. Using a multi-temperature model of the post-shock flow,
we estimate that the white dwarf in both systems has a mass greater
than 1 \Msun. The implications of this result are discussed.  We
demonstrate that the `soft X-ray excess' observed in many magnetic
cataclysmic variables can be partially attributed to using an
inappropriate model for the hard X-ray emission.
\end{abstract}


\section{Introduction}

Polars, or AM Her systems, are interacting binary systems in which the
magnetic field of the accreting white dwarf is strong enough to
prevent the formation of an accretion disc. It also forces the spin
period of the white dwarf to be synchronised with the binary orbital
period. In these systems, the accreting material leaves the red dwarf
secondary, follows a ballistic trajectory before getting coupled by
the magnetic field of the white dwarf and funnelled directly
onto the white dwarf surface.

Just above the white dwarf photosphere, the accretion flow is strongly
shocked, at which point the temperature reaches $\sim$50keV.  Some of
this radiation is intercepted by the white dwarf where it is
thermalised and re-radiated as soft X-rays. An additional source of
soft X-rays is dense blobs of material which penetrate into the
white dwarf photosphere and heat it directly.  In addition, there is
a cyclotron component which is strongest in the optical band (for the
magnetic field strength found in polars: $\sim$10--200MG). This arises
from electrons spiralling around the magnetic field lines close to the
shock.

Prior to the launch of the {\sl ROSAT} X-ray satellite in 1990 the
relative strength of the three emission components was a major issue
(eg King \& Watson 1987). Observations of several polars showed that
the strength of the soft X-ray component relative to the hard X-ray
component was in excess of that predicted by simple accretion models
(eg Heise et al 1985). With the results derived using {\sl ROSAT} it
became clear that many polars showed this `soft X-ray excess' (Ramsay
et al. 1994).  During epochs when the accretion rate was in an
`intermediate' level of accretion, Ramsay, Cropper \& Mason (1995)
found that the relative strength of the soft X-ray flux was less than
in `high' accretion states: indeed AM Her showed a `soft X-ray
deficiency' in this state. They hypothesized that for lower accretion
rates, a shock does not form over the whole of the accretion region.

A significant uncertainty in the results of Ramsay et al (1994) was
the poor constraint on the hard X-ray flux since {\sl ROSAT} did not
extend to energies much higher than $\sim$2keV. With the launch of
{\sl XMM-Newton} we are able, for the first time, to detect both the
soft and hard X-ray components clearly. This paper presents an
analysis of the spectra of the first polars to be observed using {\sl
XMM-Newton}.

\section{Observations}

DP Leo and WW Hor were observed using {\sl XMM-Newton} in Nov and Dec
2000 respectively: the observation log is shown in Table
\ref{obslog}. Both objects were detected in all three EPIC detectors
(Turner et al 2001, St\"{u}der et al 2001), which were configured in
full window mode and used the thin filter. Neither object was detected
in the RGS detectors. During the DP Leo observations, there were time
intervals of enhanced particle background. These intervals were
removed from the spectral analysis, reducing the exposure by 16ksec in
the EPIC detectors. The observations of WW Hor were not significantly
effected by increased particle background.  By comparing the X-ray and
optical flux with previous observations, it was concluded that both
systems were in an intermediate state of accretion (cf Pandel et al
2001).

Before extracting source spectra, the data were processed using the
first public release of the {\sl XMM-Newton} Science Analysis System.
Spectra were extracted from the EPIC PN camera data using an aperture
of $\sim19^{''}$ in radius centered on the source, chosen so that the
aperture did not cover more than one CCD. This encompasses $\sim$85
percent of the integrated PSF (Aschenbach et al 2000). In the case of
the EPIC MOS data, the source was in the centre of the CCD and a
radius of $\sim24^{''}$ was used (encompassing $\sim$90 percent of the
integrated PSF). Background spectra were extracted from the same CCD
on which the source was detected, scaled and subtracted from the
source spectra.

Since the response of the EPIC PN detector is not well calibrated at
present below $\sim$0.2keV, energies below this were ignored in the
subsequent analysis. Also the PN response function used assumes only
single pixel events, so we extracted only these events to construct our
PN spectrum.  For the EPIC MOS detectors, which are currently better
calibrated at lower energies, we fitted energies above 0.1keV. The
response files used in the analysis are listed in Table
\ref{response}.

\begin{table}
\begin{tabular}{llr}
Object&Observation&Effective exposure  (ksec)\\
      &Date &  PN, MOS\\
\hline
DP Leo& 22 Nov 2000& 19, 18\\
WW Hor& 4 Dec 2000& 21, 23\\
\hline
\end{tabular}
\caption{The observation log of the {\sl XMM-Newton}
observations of DP Leo and WW Hor. In the case of the DP Leo
observations, time intervals of enhanced particle background were
excluded.}
\label{obslog}
\end{table}

\begin{table}
\begin{tabular}{lr}
EPIC detector & response file used\\
\hline
PN & epn$\_$fs20$\_$Y9$\_$thin.rmf\\
MOS1 & m1$\_$thin1v9q19t5r4$\_$all$\_$15.rsp\\
MOS2 & m2$\_$thin1v9q19t5r4$\_$all$\_$15.rsp\\
\hline
\end{tabular}
\caption{The response files used in the analysis.}
\label{response}
\end{table}

\section{The model for the X-ray spectrum}
\label{model}

The hard X-ray component produced in the hot post-shock flow has
traditionally been modelled using a single temperature thermal
bremsstrahlung component. We have developed a physically more
realistic model of the post-shock flow (Wu, Chanmugam \& Shaviv 1994,
Cropper, Ramsay \& Wu 1998), which takes account of its
multi-temperature nature and the fact that a significant fraction of
the cooling takes place in the form of cyclotron radiation. It also
includes the hard X-rays reflected from the surface of the white
dwarf. In addition, Cropper et al (1999) included the effect of the
varying gravitational force over the height of the post-shock flow. In
this paper we use the model of Cropper et al (1999)
to model the hard X-ray component and a blackbody to model any soft
X-ray component. In addition, we account for absorption using a
neutral absorption model.

\section{DP Leo}
\label{dpleo}

\begin{figure*}
\begin{center}
\setlength{\unitlength}{1cm}
\begin{picture}(15,10)
\put(-16,-1.){\includegraphics{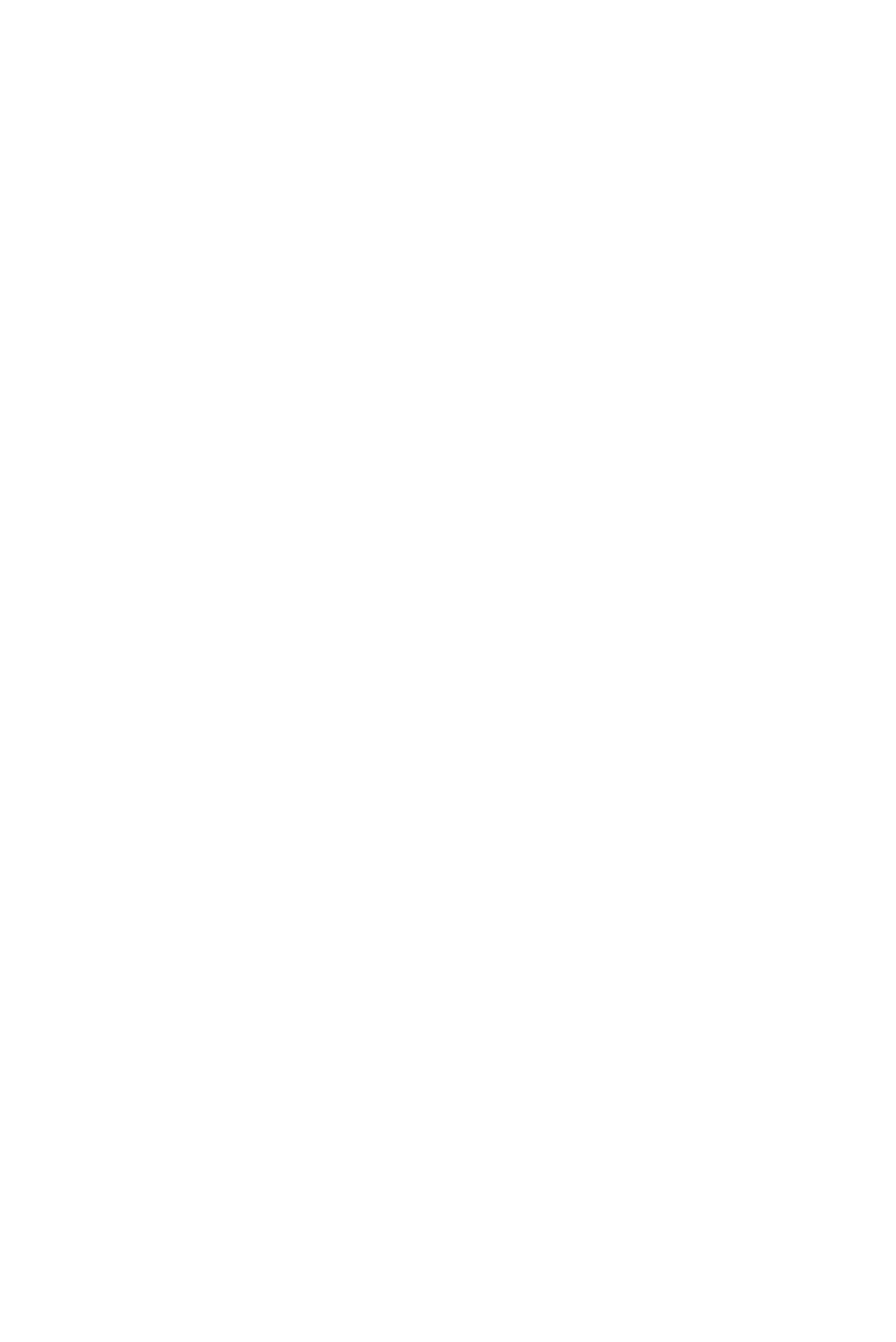}}
\put(-4.2,3.8){\includegraphics{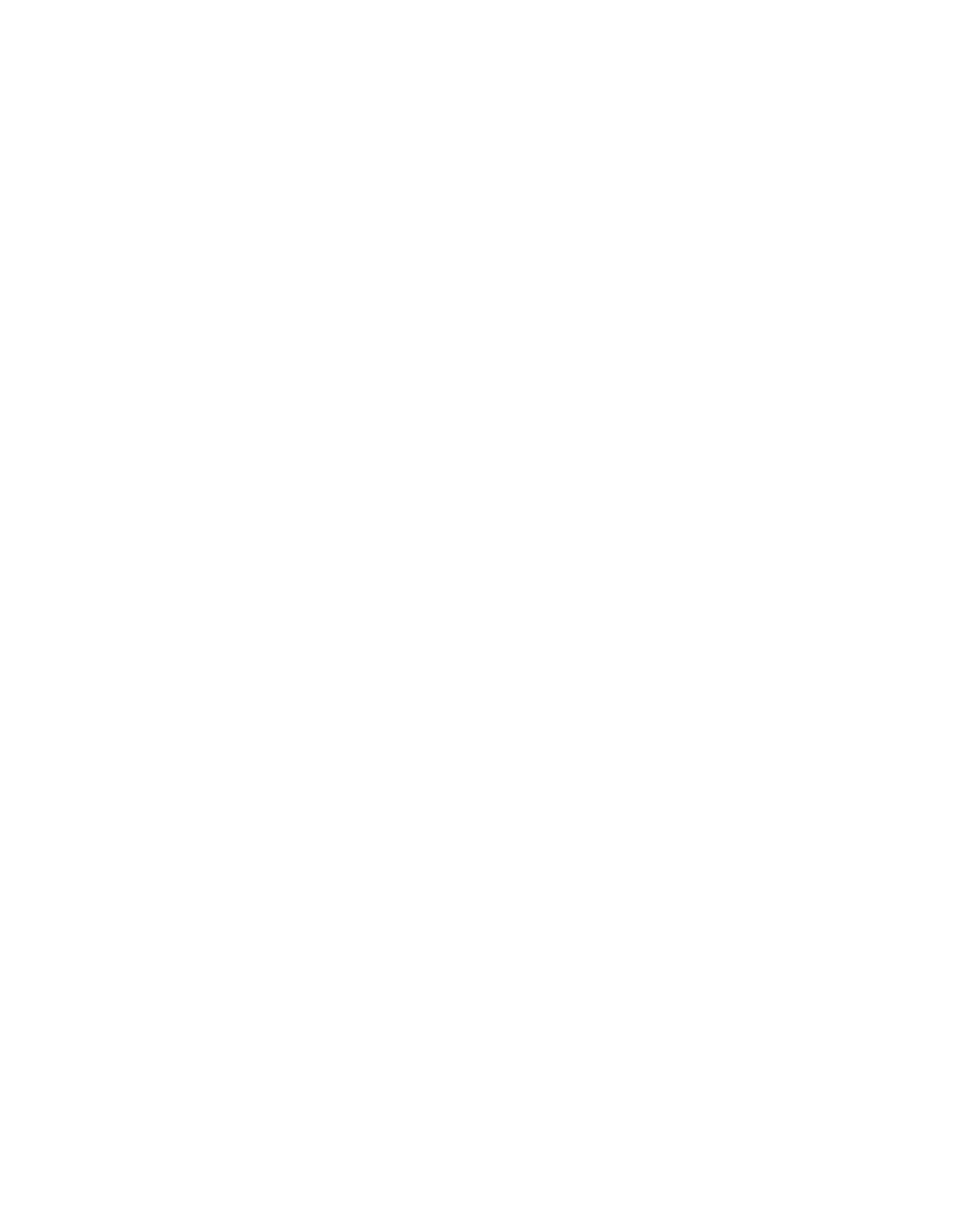}}
\end{picture}
\end{center}
\caption{The EPIC PN (upper), MOS1 (lower) X-ray spectra of DP
Leo. The solid lines show the best fit using an absorbed blackbody
plus multi-temperature accretion shock model.}
\label{dpspec} 
\end{figure*}

The integrated EPIC PN and MOS1 spectra of DP Leo are shown in Figure
\ref{dpspec}. A soft X-ray component is seen together with a hard
component -- this is the first time that DP Leo has been detected
above 2keV. Line emission from Fe K$\alpha$ is detected at
$\sim$6.7keV.

We fitted all three EPIC X-ray spectra separately using the model
described in \S \ref{model}. We fix the specific accretion rate,
\mdot, at 1 g s$^{-1}$ cm$^{-2}$ (appropriate for a system in an
intermediate accretion state), the ratio of cyclotron to
bremsstrahlung cooling, $\epsilon_{s}$, at 10 (appropriate for the
magnetic field strength seen in DP Leo: 31MG Cropper \& Wickramasinghe
1993) and the metal abundance at solar. The fit is not sensitive to
the exact values chosen for these parameters.

The best fit parameters are shown in Table \ref{fits}: good fits are
achieved. The absorption towards DP Leo is low, $<10^{20}$ cm$^{-2}$
and the temperature of the blackbody is similar to that found for
other polars (eg Ramsay et al 1994). To determine the implied mass of
the white dwarf we assume the Nauenberg (1972) mass-radius
relationship for white dwarfs. This yields a white dwarf mass
greater than 1.3\Msun.

We also show in Figure 1, the spectra centered on 7keV. Our model fits
underestimate the line flux for the Fe K$\alpha$ emission lines.  To
investigate this further, we varied the metal abundance,
$\epsilon_{s}$ and \mdot.  No solution was achievable
that was consistent with both the line and continuum flux. We address this
further in \S \ref{discuss}.

We show in Table \ref{fits} the unabsorbed, bolometric fluxes for the
soft and hard X-ray components. The soft X-ray luminosity is defined
as $L_{soft,bol}=\pi f_{soft,bol} \sec(i-\beta) d^{2}$, where we
assume that the soft X-ray emission is optically thick and can be
approximated by a small thin slab of material, the unabsorbed
bolometric blackbody flux is $f_{soft,bol}$, $d$ is the distance, $i$
is the system inclination and $\beta$ is the angle between the
accretion region and the spin axis. For DP Leo, $i=80^{\circ}$ and
$\beta=100^{\circ}$ (Bailey et al 1993), while Biermann et al (1985)
find $d > 380$ pc.  Because we have been able to extend to lower
energies in the MOS detectors compared to the PN detector (since they
are currently better calibrated), the absorption is more constrained
using the MOS detectors. In determining the flux from the soft
component using the PN data we have constrained the absorption to be
$<9\times10^{19}$ cm$^{-2}$ -- the upper limit derived from the MOS
detectors.

We define the hard X-ray luminosity as $L_{hard,bol}=4\pi f_{hard,bol}
d^{2}$, where $f_{hard,bol}$ is the unabsorbed, bolometric flux from
the hard component. Since a fraction of the hard X-ray flux is
directed towards the white dwarf and some of that is reflected back
towards the observer, we switch the reflected component to zero after
the final fit to determine the intrinsic flux from the optically thin
shocked material. 

We also show in Table \ref{fits} our estimate of the ratio
$L_{soft,bol}/L_{hard,bol}$. Using data obtained using {\sl ROSAT},
Ramsay et al (1994) were able to put an upper limit only on this ratio
since the hard X-ray component was not significantly detected. Now
using the {\sl XMM-Newton} data, we can place relatively good
constraints on the ratio because we sample both the hard and soft
X-ray components. If there were no dense blobs of material in the
accretion stream, we expect $L_{soft,bol}/L_{hard,bol}\sim0.5$ (in
some polars in a high state of accretion, it is found that this ratio
is much greater than this; on the other hand, for systems in
intermediate states of accretion, the ratio can be much lower). The
$L_{soft,bol}/L_{hard,bol}$ ratio in DP~Leo, is around unity, implying
consistency with the standard Lamb and Masters (1979) model.

Cropper, Wu \& Ramsay (2000) suggest that using a single temperature
thermal bremsstrahlung to model the hard X-ray component will
contribute to the apparent `soft X-ray excess' seen in many
polars. This is because a fraction of the photons attributed to the
soft X-ray component actually originates in the lowest levels of the
post-shock flow rather than from re-processing by the white dwarf. Our
multi-temperature shock model intrinsically emits a significant amount
of soft X-rays at the base of the post-shock flow. It is interesting
to determine the soft to hard X-ray ratio if we assume an absorbed
blackbody plus single temperature thermal bremsstrahlung model as used
by Ramsay et al (1994). Fixing the thermal bremsstrahlung temperature
at 30keV (as there), and adding a Gaussian to model the Fe K$\alpha$
line emission at $\sim$6.7keV, we find that the ratio,
$L_{soft,bol}/L_{hard,bol}$ is twice the value determined using the
multi-temperature model. This suggests that when comparing the
relative strength of the shocked and re-processed X-ray components, it
is essential to use the most physically realistic model of the
post-shock flow available.

\small 
\begin{table*}
\begin{tabular}{llrrrrrrrrr}
& & $N_{H}$ & $kT_{bb}$ & $M_{1}$ &Flux$_{soft,bol}$ &
$L_{soft,bol}$ & Flux$_{hard,bol}$ & $L_{hard,bol}$ & $L_{soft,bol}/$& \rchi\\ 
& & ($\times10^{19}$ & (eV) & (\Msun) & ($\times10^{-12}$ erg 
& ($\times 10^{31}$& ($\times10^{-13}$ erg& ($\times
10^{31}$ &$L_{hard,bol}$ & (dof)\\ 
& & cm$^{-2}$) & & & s$^{-1}$ cm$^{-2}$) & \ergss)& s$^{-1}$ cm$^{-2}$) & 
\ergss) & & \\
\hline 
DP Leo& PN &0.0$^{+20}$ & 23$^{+4}_{-5}$ & 1.40$_{-0.12}$ & $6.7^{+36.2}_{-3.7}$ & 3.4$^{+18.6}_{-1.4}$&
$4.4^{+0.8}_{-0.5}$ & $0.8^{+0.1}_{-0.1}$ & $4.2^{+29}_{-3.0}$& 
1.12 (126) \\
 & MOS1 &$2.1^{+6.9}_{-0.6}$ & $30^{+3}_{-6}$ & 1.40$_{-0.09}$& 
2.5$^{+9.5}_{-0.8}$ &
1.3$^{+4.9}_{-0.5}$ & 6.6$^{+1.0}_{-1.0}$ & 1.3$^{+0.1}_{-0.2}$ & 
1.0$^{+5.7}_{-0.5}$ &0.88 (90)\\
 & MOS2 & 3.3$^{+5.7}_{-1.8}$ & 28$^{+3}_{-5}$ & 1.40$_{-0.08}$& 
3.3$^{+3.7}_{-1.5}$& 1.6$^{+2.0}_{-0.7}$ & 6.6$^{+1.3}_{-0.9}$ &
1.3$^{+0.2}_{-0.1}$ & 1.2$^{+2.4}_{-0.8}$ & 1.05 (72) \\
\hline
WW Hor & PN & 0.0$^{+1.4}$ & & 1.04$^{+0.05}_{-0.06}$& & &
7.6$^{+0.4}_{-0.3}$ & 1.7$^{+0.1}_{-0.1}$ & & 1.25 (123)\\  
 & MOS1 &1.4$^{+4.2}_{-1.2}$ & & 1.18$^{+0.10}_{-0.09}$ & & &
12.8$^{+1.0}_{-1.5}$ & 2.8$^{+0.2}_{-0.2}$ & & 1.02 (116)\\
 & MOS2& 0.0$^{+1.6}$ & & 0.99$^{+0.06}_{-0.05}$ & & &
7.6$^{+0.4}_{-0.5}$ & 1.7$^{+0.1}_{-0.1}$ & & 0.98 (165)\\
\end{tabular}
\caption{The best fit parameters to the EPIC PN, 
MOS1 and MOS2 data of DP Leo and WW Hor using the multi-temperature
model described in \S \ref{model}. A distance of 400pc is assumed
for DP Leo (Biermann et al 1985 find $d>380$pc) and 430pc for WW Hor
(Bailey et al 1998). In determining the fluxes and luminosities
of DP Leo using the PN data, we have chosen an upper limit of
$N_{H}<9\times10^{19}$ cm$^{-2}$ which is the upper limit derived from
the MOS data (which has a better response at lower energies).}
\label{fits}
\end{table*}
\normalsize

\section{WW Hor}
\label{wwhor}

\begin{figure*}
\begin{center}
\setlength{\unitlength}{1cm}
\begin{picture}(15,10)
\put(-16,-1.){\includegraphics{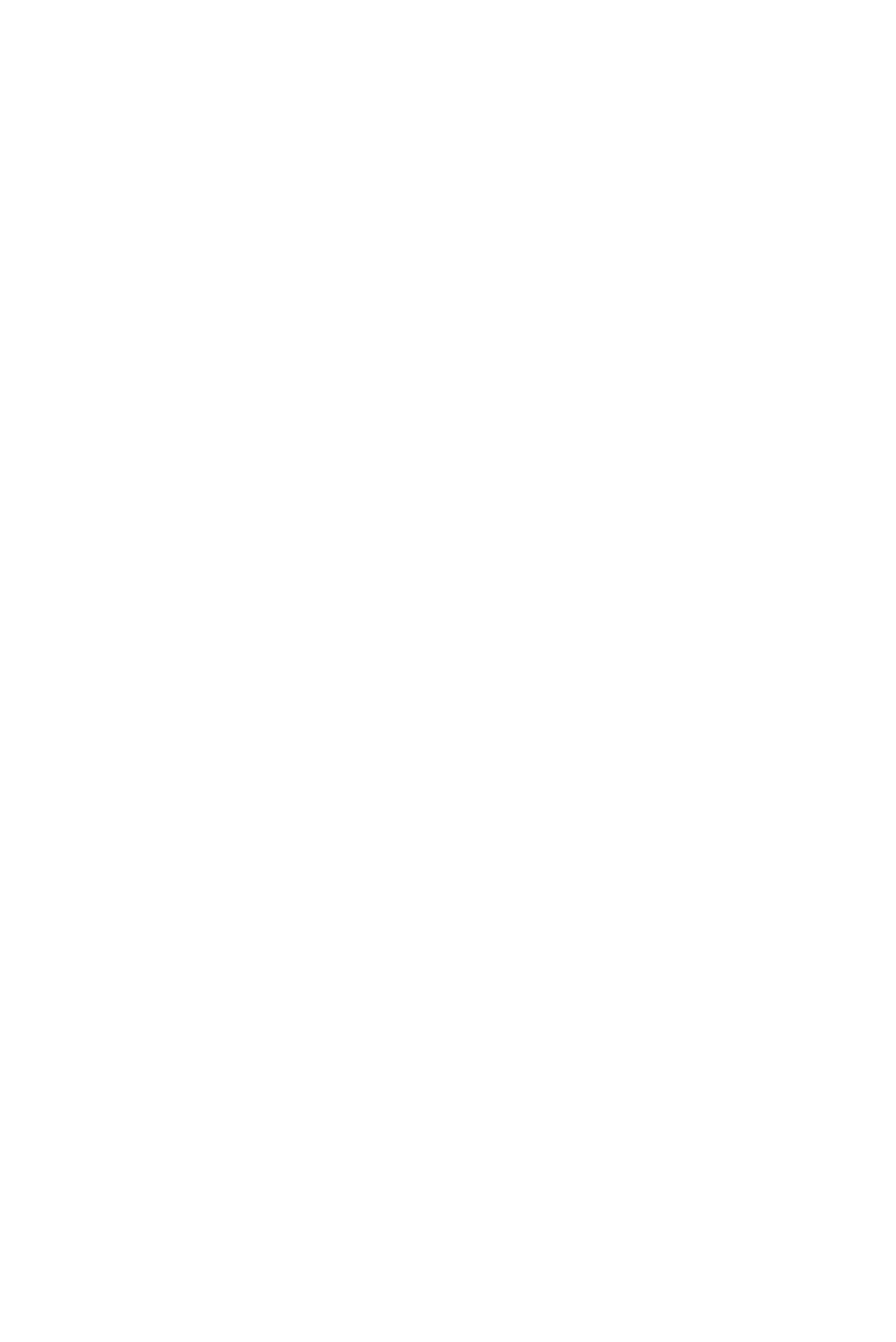}}
\put(-4,3.8){\includegraphics{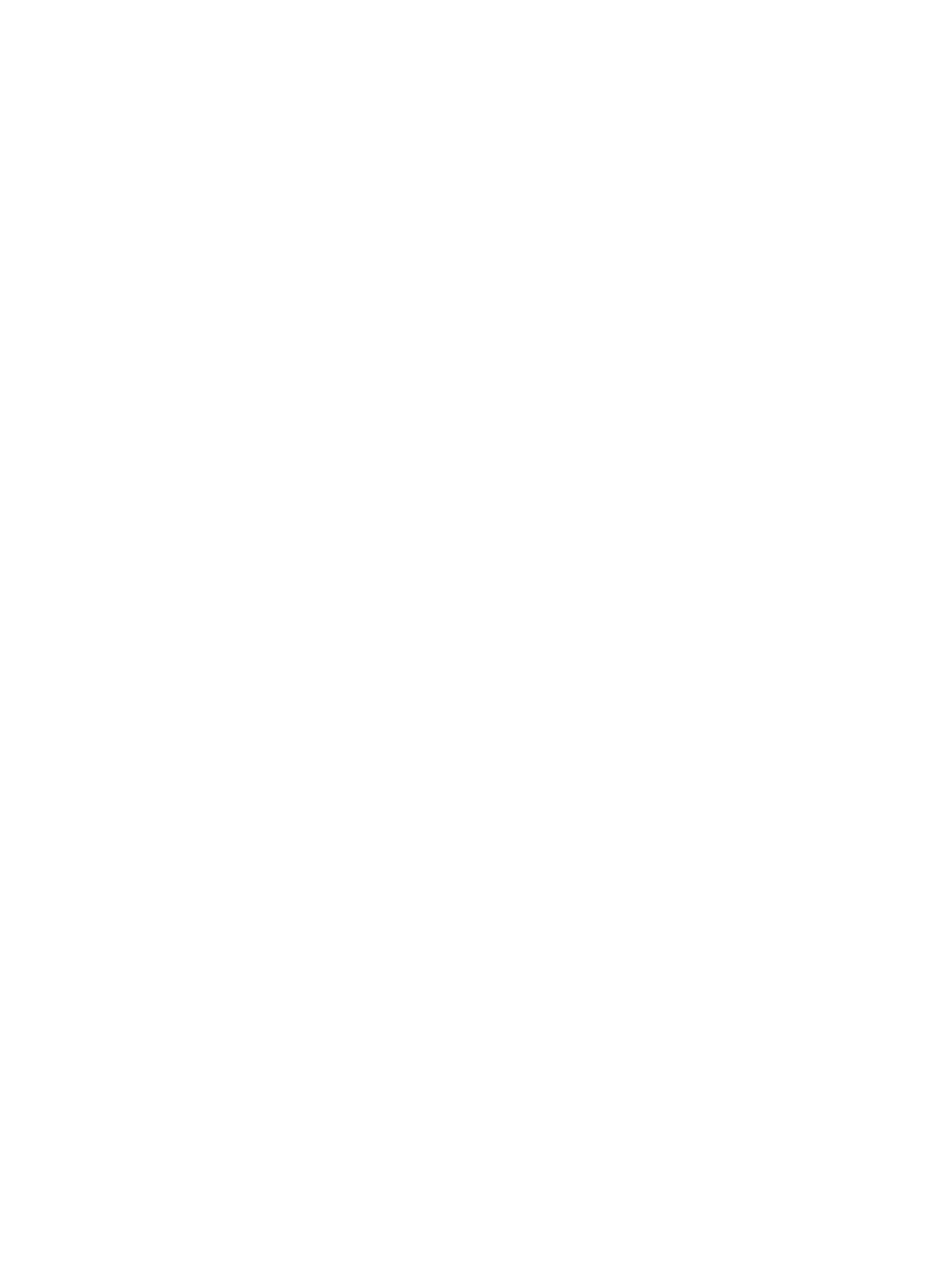}}
\end{picture}
\end{center}
\caption{As for figure 1, but for WW Hor}
\label{wwspec} 
\end{figure*}

The integrated EPIC PN and MOS1 spectra of WW Hor are shown in Figure
\ref{wwspec}. Whilst we observe hard X-ray emission above 2keV in WW
Hor for the first time, there is no evidence for a soft blackbody
component with the sort of temperature seen in DP Leo. This is
unusual, since even for systems in a reduced rate of accretion a soft
X-ray component is expected. Weak Fe K$\alpha$ line emission is
present at 6.7 keV.  In the PN spectrum there is evidence for an
emission line due to Fe fluorescence near 6.4keV.

We modelled the integrated EPIC spectra from each detector separately
using the  model described in \S \ref{model}
together with a neutral absorber. As in DP Leo we fix \mdot, at 1 g
s$^{-1}$ cm$^{-2}$, $\epsilon_{s}$ at 10 and the metal abundance at
solar. We obtain good fits: the best fit model parameters are shown in
Table \ref{fits}. We find a very low absorption column. The mass of
the white dwarf is $\sim$1.1\Msun. Using a distance of 430pc (Bailey
et al 1988) we estimate the unabsorbed, hard bolometric luminosity to
be $L_{hard}=1-2\times10^{31}$ \ergss. This is similar to that found
for other AM Her systems in intermediate accretion states (Ramsay,
Cropper \& Mason 1995). Indeed, when we compare the phase averaged
observed flux in the 0.1--2.0keV band in {\sl XMM-Newton}
($2.1\times10^{-13}$ \ergscm in MOS1) with the {\sl ROSAT} observation
of WW Hor, made when it was in an intermediate state
($2.3\times10^{-13}$ \ergscm), we find that they are very similar.

We can investigate whether a low temperature soft X-ray component
could be absorbed by a modest absorption column. In the lowest energy
bands the best calibrated instrument is the EPIC MOS1 detector. We
added a blackbody component with temperature $kT_{bb}$=10 and 15 eV to
the model used above (cf $kT_{bb}\sim$30 eV for DP Leo). 
We find that such a low temperature blackbody is
`hidden' by an absorption column consistent with our upper limit
(5.6$\times10^{19}$ \pcmsq). The corresponding values of $L_{soft}$ is
8.5$\times10^{30}$ and 4.3$\times10^{30}$ \ergss for 10 and 15eV
respectively. This corresponds to $L_{soft}/L_{hard}$=0.4 and 0.2:
these values are broadly similar to that found for polars in
intermediate accretion states (Ramsay, Cropper \& Mason 1995).

\section{Discussion}
\label{discuss}

When WW Hor was observed using {\sl ROSAT} it was found that both a
blackbody and a bremsstrahlung component were required to fit the
X-ray spectrum (Ramsay, Cropper \& Mason 1995). When WW Hor was
observed using {\sl XMM-Newton}, it was at very similar X-ray flux
levels compared to the {\sl ROSAT} observation. Further, the X-ray
bolometric luminosity of WW Hor is approximately similar to DP Leo
(implying similar mass transfer rates) if we assume similar distances.
Why do we therefore observe a soft X-ray component in DP Leo and not
WW Hor?

We suggest that the accretion region was more spread out in the {\sl
XMM-Newton} observation of WW Hor compared to the {\sl ROSAT}
observation, thereby decreasing the local heating rate and causing the
temperature of any soft X-ray component to decrease. If it was
sufficiently low (say $kT$=10 or 15eV), we showed in \S \ref{wwhor}
that even a very modest amount of absorption would prevent us from
detecting this soft X-ray component. It is not clear why the accretion
region would cover a larger area of the white dwarf. One possible
reason could be that for relatively low mass transfer rates, accretion
occurs onto a wide range of magnetic azimuths, resulting in a more
extended accretion region. This is thought to be one of the reasons
why intermediate polars (which have less strong magnetic fields than
polars) do not show a soft X-ray component (eg King \& Lasota 1990).

We now briefly comment on the derived masses for the white dwarfs in
these systems. In the case of WW Hor, we find a mass of $\sim$1.1\Msun
and for DP Leo $>$1.3\Msun (table \ref{fits}). The mass of the white
dwarf in both systems is at the upper end of expectations.  There are
no other accurate estimates for the mass of the white dwarf in these
two systems. Several magnetic CVs were found to have similar masses to
WW Hor in the survey of such systems using {\sl RXTE} data and the
same emission model as used here (Ramsay 2000). Similarly, the mass of
Sirius B is 1.034$\pm$0.026 (Holberg et al 1998). Other massive
non-accreting white dwarfs have been found (eg RE J0317--858;
1.31--1.37\Msun, Ferrario et al 1997). In the case of DP Leo, however,
the mass of the white dwarf is very heavy indeed and would be the
heaviest white dwarf within an accreting magnetic system yet found.
Ramsay (2000) found that the mass distribution of isolated
non-magnetic white dwarfs and white dwarfs in magnetic CVs are
significantly different: the white dwarfs in magnetic CVs are biased
towards heavier masses. However, it was not clear if this was due to
selection effects: it is possible that high mass polars are more
likely to be discovered than low mass polars.

In calculating the mass of the white dwarf we use the Nauenberg (1972)
approximation for the white dwarf mass-radius relationship: this
approximates the Hamada \& Salpeter (1961) mass-radius relationship
for carbon white dwarfs over the range 0.3--1.2 \Msun, and is
consistent with the best mass-radius data for the white dwarfs 40 EriB
and Sirius B (Provencal et al 1988). However the appropriate
mass-radius relationship is still a matter of debate. In the case of
DP Leo, where we estimate a white dwarf of mass $>$1.3\Msun, there is
some additional uncertainty for our mass estimate since for the very
heaviest white dwarfs the mass-radius relationship is very uncertain.
Hamada \& Salpeter (1961) also show the mass-radius relationship for
Fe core white dwarfs. White dwarfs with Fe cores are thought to be
formed when a white dwarf with a ONeMg core ignites through an
accretion-drived collapse (Isern, Canal \& Labay 1991). The maximum
mass for such a white dwarf is $\sim$1.1\Msun. Using this mass-radius
relationship we obtain a mass of $\sim$1.1\Msun, again close to the
maximum.

Alternatively, it is possible that additional cooling processes not
included in our model could become important for white dwarfs greater
than 1.0\Msun. This would have the effect of modifying the
temperature-height profile of the post-shock region and hence the
implied X-ray spectrum. If the volume of gas emitting \ltae10keV was
underestimated, then this may also account for the deficit of Fe line
emission in DP Leo (\S \ref{dpleo}). In general, however, additional
cooling terms work to soften the X-ray spectrum, resulting in a higher
gravitational potential (and hence white dwarf mass) to fit the
observed spectrum. It is also possible that the absorption is more
complex than a simple neutral absorber, as was found to be the case in
the polar BY Cam (Done \& Magdziarz 1998). This would make the
continuum appear harder, so overestimating the shock temperature and
hence the white dwarf mass. The higher temperature continuum would
also underpredict the plasma iron lines fluxes. Given that both these
features are seen in our model fits, we added a partial covering
component and also an ionized absorption component to our model. The
resulting fits and masses were not significantly improved with the
exception of the MOS1 data which gave masses which were 0.1--0.2 \Msun
lighter. Data of higher signal to noise would be required to test the
necessity of more complex absorption models.

\section{Acknowledgements}
Based on observations
obtained with XMM-Newton, an ESA science mission with instruments
and contributions directly funded by ESA Member States and the USA
(NASA).

\end{document}